\documentclass[a4paper,10pt]{article}
\usepackage{graphicx}
\usepackage{epstopdf}
\usepackage{amsfonts,amsmath,amssymb}
\usepackage{hyperref}

\usepackage{epsfig,multicol}

\newcommand\fverb{\setbox\pippobox=\hbox\bgroup\verb}
\newcommand\fverbit{\egroup\item[\fbox{\unhbox\pippobox}]}

\newbox\pippobox

\begin{document}
\title{Inflation with Gauss-Bonnet Correction and Higgs Potential}
\author{Zahra Ahghari\thanks{email: z\_ahghari@sbu.ac.ir}${}\ $ and Mehrdad Farhoudi\thanks{email:
       m-farhoudi@sbu.ac.ir}\\
 {\small Department of Physics,
         Shahid Beheshti University, 1983969411, Tehran, Iran}}

\date{May 28, 2026}
\maketitle
\begin{abstract}
\noindent
 We investigate the cosmological inflation for the Einstein-Hilbert
action plus the Higgs potential function and the Gauss-Bonnet term
coupled with the Higgs scalar field through a dilaton-like
coupling. Then, using the Friedmann-Lema\^{i}tre-Robertson-Walker
metric and considering the appropriate slow-roll parameters, we
derive the necessary equations of motion. In the proposed model,
since the $e$-folding integral cannot be easily solved
analytically, we first utilize a well-known Taylor expansion.
Then, with a certain range of values derived for the model
parameters, utilizing several plots and numerical analysis
methods, we obtain results for the tensor-to-scalar ratio and the
scalar spectral index that are in good agreement with the latest
observational data, particularly from ACT DR6, within the
acceptable range of the $e$-folding values. Meanwhile, a key
aspect of this work, crucial for achieving reliable values of the
inflationary observables, lies in the adopted functional forms for
the potential and coupling functions. Also, in the absence of the
Gauss-Bonnet term, we find that the inflationary observables are
roughly the same as the predictions of the chaotic inflation
model.
\end{abstract}

\section{Introduction}
\indent

The cosmic inflation theory has been agreed upon by many
cosmologists as a suitable way to explain fundamental problems of
the standard model of cosmology, such as the flatness problem and
the horizon problem in the early universe. For this purpose, the
inflation generated based on a canonical scalar field with minimal
coupling is one of the simplest methods available, wherein the
scalar field slowly rolls during the inflation. In this type of
model, the scalar field is called inflaton because it is the main
cause of inflation~\cite{reff1}--\cite{Kallosh}. According to this
model, the quantum fluctuations of the inflaton scalar field
affect the curvature perturbations, and with the growth of these
perturbations, cosmic structures such as galaxies and stars are
formed over time. To check the correctness of such models, the
measurement of temperature anisotropy in the cosmic microwave
background (CMB) can be used. The latest CMB data obtained from
the Planck satellite and the recent sixth data release (DR6) of
the Atacama Cosmology Telescope (ACT) collaboration show serious
limitations on the values of the scalar spectral index of
curvature perturbations, $n_{\rm S}^{}$, and the tensor-to-scalar
ratio, $r$. Such limitations indicate that there is~not much
freedom in choosing the parameters of the subject, including the
type of inflaton and its potential~\cite{slowroll7}--\cite{reff9}.

In addition to such a simple model for inflation, other relatively
complex models have also been presented by researchers to answer
the issues and problems of this field. Although the inflation
theory provides a consistent answer to the old problems of
cosmology, despite the efforts made, new questions (along with
ambiguities) have been raised, including the nature of the
inflaton scalar field and the nature of dark matter and dark
energy~\cite{reff10}.

At present, the only scalar field available in the set of
fundamental particles is the field corresponding to the Higgs
boson, which appears to be a suitable candidate for the inflaton,
see, e.g., Refs.~\cite{Kallosh,reff11}--\cite{Karananas}. In
particular, with the discovery of Higgs features in the large
hadron collider, which shows that the Higgs boson model has
significant uncertainties in its parameters compared to some of
the models used to develop the standard model. However, if the
Higgs scalar field enters the inflation directly, it will be
incompatible with the standard model of particle physics. To solve
this problem, it has been proposed to add a non-minimal coupling
to the gravity term, which has been studied the most, see, e.g.,
Refs.~\cite{reff11, reff12, Germani2010, reff13}. By considering
the non-minimal coupling between gravity and the Higgs scalar
field, it is possible to remedy the aforementioned inconsistency
problem and stabilize inflation, but new problems are encountered
at the same time. Indeed, the price given for the desirability of
this model is the unusual magnitude of the non-minimal coupling,
approximately $10^{4} $. Such a (meta-)stability of the vacuum in
the Higgs inflation model is still under discussion~\cite{reff14,
reff15}. On the other hand, due to the presence of the coupling
term with a large coupling constant, the unitarity bound of the
theory is violated because the energy scale in this inflation
model is larger than or close to the cutoff value~\cite{reff13}.
Recently, a Bayesian analysis of the Starobinsky and Higgs
inflationary models has been performed in Ref.~\cite{Zharov}.

On the other hand, in the discussion of Higgs inflation, there is
some freedom in the coupling of the Higgs scalar field with
gravity, since measurements of the large hadron collider determine
the coupling constant of the Higgs scalar field, but not~its
coupling constant with gravity. Hence, any arbitrary coupling
constant with gravity can be used for it, see, e.g.,
Ref.~\cite{reff16}. In this work, we utilize this freedom of the
Higgs scalar field and attempt to couple it with the Gauss-Bonnet
term instead of the Ricci scalar.

The Gauss-Bonnet term is the second term of the Lovelock
Lagrangian. It is also induced from the superstring theory and a
subclass of the Horndeski
theory~\cite{Horndeski74}--\cite{Horndeski2024}. However,
including the Gauss-Bonnet term alone in a four-dimensional
Lagrangian of gravitation does~not affect the equations of motion,
see, e.g., Refs.~\cite{reff17}--\cite{reff19}. Actually, it is
topologically invariant in $4$-dimensions, and hence non-dynamical
in $4$-dimensions. Nonetheless, its effect can appear when coupled
non-minimally with a scalar field function, which can then play as
an effective potential in causing inflation. Of course, it creates
violent negative instabilities for tensor perturbations around a
de~Sitter background on small
scales~\cite{jiang8}--\cite{reff10,reff20}--\cite{reff22}.

The inflation resulting from the coupling with the Gauss-Bonnet
term has been studied in
Refs.~\cite{slowroll7}--\cite{reff10,reff15,reff16,reff20}--\cite{Yogesh}.
For simplicity in calculations and analysis, in a number of
references such as~\cite{slowroll7,jiang8, reff41}, the potential
function and the coupling function ($ V(\phi)$ and $\xi(\phi)$ as
functions of the inflaton scalar field) are assumed so that the
relaxation condition $ V(\phi)\xi(\phi)$ is a constant. In these
references, the focus is on two specific models, namely
($V(\phi)=V_{0}e^{-\lambda\phi} $ and
$\xi(\phi)=\xi_{0}e^{\lambda\phi} $ with $V_{0}$ and $\xi_{0}$ as
constants) and ($ V(\phi)=V_{0}\phi^{n}$ and $
\xi(\phi)=\xi_{0}\phi^{-n} $), wherein the results of calculations
lead to the reduction of the tensor-to-scalar ratio.

In this study, we focus on breaking such a relaxation condition to
check whether the obtained results are consistent with the
observational data. Moreover, we consider the inflaton scalar
field to be a Higgs scalar field with a Higgs potential. However
within inflation, since the field is much larger than the expected
vacuum value, the potential can be approximated to a quartic form,
$ V_{0}\phi^{4} $. On the other hand, in this work, we consider
the coupling function as dilaton-like, i.e., $
\xi(\phi)=\xi_{0}e^{-\lambda\phi} $, and investigate the scalar
spectral index and the tensor-to-scalar ratio, and compare the
results with the recent constraints by the Planck
collaboration~\cite{Planck2018VI,Planck2018X}
\begin{align}\label{planckdata}
&n_{\rm S}= 0.9649\pm 0.0042\quad \nonumber\\
&{}\qquad\left({\rm at}\ 68 \%\, {\rm CL, Planck TT,TE,EE+lowE+lensing}\right)\!,\nonumber\\
& r<0.10 \quad\left({\rm at}\ 95 \%\, {\rm CL, Planck
TT+lowE+lensing}\right)\!,
\end{align}
the joint Planck, BK$15$ and BAO data, which places a further
constraint on the upper limit of $r$,
\begin{equation}\label{planckdata2}
r<0.056\quad{\rm at}\ 95 \%\, {\rm CL},
\end{equation}
and the most recent P-ACT-LB DR6~\cite{ACTcollab1,ACTcollab2}
data, which reports a higher $n_{\rm S}^{}$ value and still
tightens the upper limit of $r$,
\begin{eqnarray}\label{ACTdata}
 &&n_{\rm S}= 0.9743\pm 0.0034,\cr
 &&r<0.038\quad{\rm at}\ 95 \%\, {\rm CL}.
\end{eqnarray}

Also, as another phenomenological aspect of the Gauss-Bonnet
correction, we investigate the issue of gravitational waves, see,
e.g., Ref.~\cite{kawaiGW}. In this regard, by directly measuring
the gravitational waves resulting from the merger of neutron star
GW$170817$~\cite{LIGO1} and comparing them to their
electromagnetic counterpart GRB $170817$A~\cite{LIGO2}, the
observable range of the speed of gravitational waves, $c_{_{\rm
GW}}(=c_{\rm T}^{})$, has been determined to be
\begin{equation}
-3\times10^{-15}\leq c_{_{\rm GW}}/c -1\leq 7\times 10^{-16}
\end{equation}
and equivalently
\begin{equation}\label{GWdata1}
-6\times 10^{-15}< c_{_{\rm GW}}^{2}/c^2 -1\leq 1.4\times
10^{-15},
\end{equation}
where $c$ denotes the speed of light and $c_{\rm T}^{}$ shows the
speed of tensor perturbation modes. As is evident, in reality,
this order of difference is negligible to a certain extent.

In this regard, the work is organized as follows. In the next
section, we first introduce the proposed action, which includes
the Gauss-Bonnet term with a coupled scalar field. Then, we derive
the modified Friedmann equations in the framework of
Friedmann-Lema\^{i}tre-Robertson-Walker (FLRW) and review the
corresponding slow-roll inflation with the necessary equations. In
Sec.~III, we introduce the proposed model and proceed the
calculations while considering the Higgs potential and a
dilaton-like coupling function. In this section, we use several
plots and numerical analysis methods to get the best model
parameters to determine acceptable inflationary observables. We
also obtain the inflationary observables in the absence of the
Gauss-Bonnet term. In addition, we consider the speed of
gravitational waves for the proposed model during inflation.
Finally, in Sec.~IV, we summarize the concluding results obtained.

\section{Slow-Roll Inflation with Gauss-Bonnet Correction}
\indent

We consider the following action in a four-dimensional spacetime
as\footnote{Note that, the Higgs inflation action is usually
considered by the coupling function $F(\phi)=1+f\phi^2$ (with $f$
as a constant) non-minimally coupled to the Ricci scalar in the
Jordan frame. In this work, we consider $f=0$ without
mentioning/resorting the Jordan frame. However, for the issue of
the Jordan frame and/or the Einstein frame in this matter, see,
e.g., Refs.~\cite{reff15,reff41}.}
\begin{equation}\label{a1}
  S\! =\! \int\! d^{4}x   \sqrt{-g}\left[ \frac{1}{2\kappa}R-\dfrac{\omega}{2}
  \left( \nabla \phi\right)^{2}- V(\phi)-\frac{1}{2}\xi(\phi)R^{2}_{\rm GB}\right]\!,
\end{equation}
where $R$ is the Ricci scalar, $R^{2}_{\rm GB}\equiv
R_{\mu\nu\rho\sigma}R^{\mu\nu\rho\sigma}-4R_{\mu\nu}R^{\mu\nu}+R^{2}
$ is the Gauss-Bonnet term, and $\phi$ is the inflaton scalar
field with a potential $V(\phi)$ and is non-minimally coupled to
the Gauss-Bonnet term with the coupling function $\xi(\phi)$. In
this study, we consider the coupling constant, $\omega$, to be
$\omega = 1 $, and $\kappa\equiv 8\pi G/c^4 $ also to be $\kappa
=1$ in the Planckian units including the natural units,
$\hbar=1=c$.

The variation of action (\ref{a1}) with respect to the inflaton
scalar field gives the modified Klein-Gordon equation
\begin{equation}\label{a2}
\square \phi -V^{'}(\phi)-\frac{1}{2}\xi ^{'}(\phi) R^{2}_{\rm GB}
=0,
\end{equation}
where $\square ={}_{;\alpha}{}^{;\alpha}$, the prime is the
derivative with respect to $\phi$, and the lowercase Greek indices
run from zero to three. Also, the variation of action (\ref{a1})
with respect to the metric yields  the modified Einstein equation
\begin{equation}\label{a3}
\begin{split}
& (R^{\mu \nu}\!-\frac{1}{2}g^{\mu\nu}R )-\nabla^{\mu}\phi
\nabla^{\nu}\phi+g^{\mu\nu}V(\phi)
 +\frac{1}{2}g^{\mu\nu}\nabla ^{\lambda}\phi \nabla _{\lambda}\phi\\
&-\!2g^{\mu\nu}R\,\square \xi+\!2R\nabla ^{\mu}\nabla ^{\nu}\xi
 -\!4\nabla _{\lambda}\nabla ^{\mu}\xi R^{\lambda\nu}-\!4\nabla _{\lambda}\nabla^{\nu}\xi
 R^{\lambda\mu} \\
  &+4\,\square \xi R^{\mu\nu}+4 g^{\mu\nu}\nabla _{\alpha}\nabla _{\beta}\xi
 R^{\alpha\beta}-4\nabla_{\alpha}\nabla _{\beta}\xi
 R^{\mu\alpha\nu\beta}=0.
\end{split}
\end{equation}
Note that, only the terms related to the derivatives of
$\xi(\phi)$ are appeared, wherein, according to the topological
theorem, the terms with the coefficient $\xi(\phi)$ are omitted.
In the continuation, we consider the spatially flat FLRW metric
\begin{equation}\label{metric}
ds^{2}=-dt^{2}+a^{2}(t)\left(dx^{2}+dy^{2}+dz^{2}\right),
\end{equation}
where $a(t)$ is the scale factor as a function of the cosmic time
$t$. Also, by accepting the homogeneity and isotropy, we consider
the inflaton scalar field to be only a function of the cosmic
time. Then, we insert metric \eqref{metric} into Eqs. (\ref{a2})
and (\ref{a3}), and respectively get the first and second modified
Friedmann equations and the modified Klein-Gordon equation
\begin{equation}
-3H^{2}+\frac{1}{2}\dot{\phi}^{2}+V(\phi)+12H^{3}\dot{\xi} =0,
\label{a4}
\end{equation}
\begin{equation}
-H^{2}\left(1-4\ddot{\xi} \right)-2\dfrac{\ddot{a}}{a}\left(
1-4H\dot{\xi}\right)+V(\phi)-\frac{1}{2}\dot{\phi}^{2}=0,
\label{a5}
\end{equation}
\begin{equation}
12H^{2}\dfrac{\ddot{a}}{a}\xi ^{'}(\phi)+\ddot{\phi}+3H\dot{\phi}+V^{'}(\phi) =0,
\label{a6}
\end{equation}
where $H=\dot{a}/{a}$ is the Hubble parameter and in turn
obviously $\ddot{a}/a=\dot{H}+H^2$, and the dot represents the
time derivative. It is well-known that, knowing the functionality
of $V(\phi)$ and $\xi(\phi)$ in term of $\phi $, the above three
equations are actually two independent ones for the two unknowns
$a(t)$ and $\phi(t)$.

The method of slow-roll parameters is usually used to obtain the
equations of motion in the inflation. In this case, since action
(\ref{a1}) has an additional term including a scalar field
dependence, we are required to define new slow-roll parameters in
addition to the usual slow-roll parameters used with the
Einstein-Hilbert action. Such definitions should be in a way to
ensure that the slow-roll conditions are still satisfied and the
derivatives of $ \xi (\phi)$ varies slowly. According to this
requirement, the slow-roll parameters can be introduced
as~\cite{reff42}
\begin{equation}\label{a7}
\epsilon _{1}=-\dfrac{\dot{H}}{H^{2}}=-\dfrac{d\ln H}{d\ln
a},\qquad \epsilon_{n+1}=\dfrac{d\ln \mid \epsilon _{n}\mid}{d\ln
a},
\end{equation}
\begin{equation}\label{a8}
\delta_{1}=4H\dot{\xi},\qquad \delta _{n+1}=\dfrac{d\ln \mid
\delta _{n}\mid}{d\ln a},
\end{equation}
for $n\geqslant 1$ as an integer number, and where at
least\footnote{Otherwise, if the coupling function $\xi$ is
constant, the Gauss-Bonnet term will have no effect in
four-dimensional spacetime. }\
 $\dot{\xi}\neq 0$. Thereupon, the approximation
of the slow-roll parameters in the inflation are $
\mid\epsilon_{n}\mid \ll 1 $ and $ \mid\delta_{n}\mid \ll 1 $. By
imposing these conditions, one obtains
\begin{equation}\label{Conditions1,2}
\dot{\phi}^{2}\ll V(\phi), \qquad\quad
\mid\ddot{\phi}\mid\ll\,12H^{4}\mid\! \xi^{'} \!\mid,
\end{equation}
\begin{equation}\label{Conditions3,4}
4H\mid\dot{\xi}\mid  \ll 1, \qquad\quad
\mid\ddot{\xi}\mid\ll\,H\mid \dot{\xi}\mid.
\end{equation}
Accordingly, the equations of motion (\ref{a4}), (\ref{a5}) and
(\ref{a6}) can be rewritten as
\begin{equation}
H^{2}\simeq \dfrac{1}{3}V,
\label{a9}
\end{equation}
\begin{equation}
\dot{H}\simeq -\dfrac{1}{2} \dot{\phi}^{2}-2H^{3}\dot{\xi},
\label{a10}
\end{equation}
\begin{equation}
\dot{\phi}\simeq -\dfrac{1}{3 H}\left(V^{'}+12H^{4}\xi ^{'}
\right). \label{a11}
\end{equation}
Also, in this case, the $e$-folding number throughout the
inflationary epoch generally is~\cite{Liddle1999, Baumann2009}
\begin{equation}\label{a12}
N(\phi)=\int^{\phi_{\rm end}}_{\phi}
\dfrac{H}{\dot{\phi}}d\phi\simeq \int_{\phi _{\rm
end}}^{\phi}\dfrac{V}{V^{'}+4\xi^{'}V^{2}/3} d\phi,
\end{equation}
where we have utilized the approximate Eqs.
(\ref{a9})-(\ref{a11}), and the subscript `end' denotes the value
of the inflaton scalar field at the end of inflation, i.e., when
$\epsilon_1\!\!\mid_{\phi_{\rm end}}^{}\simeq 1$.

Now, to facilitate the writing of equations, the quantity
$Q(\phi)\equiv V^{'}/V+4\xi ^{'}V/3$ can be introduced. With this
quantity, and by using the approximate
Eqs.~(\ref{a9})-(\ref{a11}), the $e$-folding number and the
slow-roll parameters can be rewritten in terms of potential and
coupling function as
\begin{equation}
N(\phi)\simeq \int _{\phi_{\rm end}}^{\phi}\dfrac{d\phi}{Q},
\label{a13}
\end{equation}
\begin{equation}
\epsilon_{1}\simeq\frac{Q}{2}\frac{V^{'}}{V}, \label{a14}
\end{equation}
\begin{equation}
\epsilon _{2}\simeq-Q\left( \frac{V^{''}}{V^{'}}-\frac{V^{'}}{V}+\frac{Q^{'}}{Q} \right),
\label{a15}
\end{equation}
\begin{equation}
\delta _{1}\simeq-\frac{4Q}{3}\xi ^{'}V,
\label{a16}
\end{equation}
\begin{equation}
\delta _{2}\simeq-Q\left( \frac{\xi ^{''}}{\xi ^{'}}+\frac{V^{'}}{V}+\frac{Q^{'}}{Q} \right).
\label{a17}
\end{equation}
Hence, the specific functional forms of the potential and coupling
functions play a crucial role in such models.

Meanwhile, for later use, let us recall the following quantities.
In Refs.~\cite{slowroll7,reff15,reff43,reff44}, the primordial
power spectrum of scalar (curvature) and tensor perturbations is
derived to be
\begin{equation}
\mathcal{P}_{\rm S}=\frac{H^{2}}{4\pi ^{2}c_{\rm S}^{3}F_{\rm S}},
\label{a18}
\end{equation}
\begin{equation}
\mathcal{P}_{\rm T}=\frac{2H^{2}}{\pi ^{2}c_{\rm T}^{3}F_{\rm T}},
\label{a19}
\end{equation}
respectively for the time of horizon crossing at $c_{\rm S}^{}k=aH
$ and $c_{\rm T}^{}k=aH $ with the comoving wavenumber $k$, where
to the lowest-order in the slow-roll parameters the difference of
time of horizon crossing is unimportant~\cite{slowroll7}. Also,
the time derivative of the slow-roll parameters is neglected
during the slow-roll inflation, and the squared propagation speed
of the scalar and tensor perturbation modes and the other
quantities appearing in relations (\ref{a18}) and (\ref{a19})
are~\cite{HwangNoh,Odintsov2021}
\begin{equation}
c_{\rm S}^{2}= 1+\frac{8\dot{\xi}\left[4 \dot{H}\dot{\xi}+\Delta
H(\ddot{\xi}-H\dot{\xi})\right]}{F_{\rm T}F_{\rm S}(1-\Delta
/2)^{2}}, \label{a20}
\end{equation}
\begin{equation}
c_{\rm T}^{2}= 1-\frac{4(\ddot{\xi}-H\dot{\xi})}{F_{\rm T}},
\label{a21}
\end{equation}
where
\begin{equation}
F_{\rm S}\equiv\frac{\dot{\phi}^{2}+6\Delta H^{3}
\dot{\xi}}{(1-\Delta /2)^{2}H^{2}}, \label{a22}
\end{equation}
\begin{equation}
F_{\rm T}\equiv 1-4H\dot{\xi}, \label{a23}
\end{equation}
and $ \Delta\equiv (1-F_{\rm T})/F_{\rm T} $. Accordingly, the
tensor-to-scalar ratio $r\equiv \mathcal{P}_{\rm
T}/\mathcal{P}_{\rm S}$ and the scalar and the tensor spectral
indices can be obtained in terms of the slow-roll parameters as
\begin{equation}
r\simeq 8\mid\! 2\epsilon_{1}-\delta_{1}\!\mid, \label{a24}
\end{equation}
\begin{equation}
n_{\rm S}^{}-1\simeq -2\epsilon _{1}-\frac{2\epsilon _{1}\epsilon
_{2}-\delta _{1}\delta _{2}}{2\epsilon _{1}-\delta _{1}},
\label{a25}
\end{equation}
\begin{equation}
n_{\rm T}^{}\simeq -2\epsilon _{1}. \label{a26}
\end{equation}

\section{The Proposed Model}
\indent

We intend to investigate the above results for a model where the
inflaton scalar field is a Higgs scalar field. Also, in this
model, we consider the potential to be the Higgs field potential,
i.e., $V(\phi)=V_{0}(\phi^{2}-\nu_{\phi}^{2})^{2}$. However, since
the inflaton scalar field in inflation is much larger than the
vacuum expectation value of the Higgs scalar field, i.e., $\phi\gg
\nu_{\phi}$, the potential can be approximated as the quartic
form\footnote{The quartic self-interaction coefficient $V_0$ is
determined by the experimentally measured values of the mass of
Higgs boson and its vacuum expectation value. In this case, the
value of $V_0$ is approximately of the order of ${\cal
O}(10^{-1})$, see, e.g., Ref.~\cite{Navas}. However, if needed as
an inflaton scalar field, its value can be approximately of the
order of ${\cal O}(10^{-13})$, see, e.g., Ref.~\cite{reff2}.}
\begin{equation}
 V(\phi)\simeq V_{0}\phi^{4},
\label{pot}
\end{equation}
where $V_0$ is a non-zero constant. Moreover, in this model, we
consider the Gauss-Bonnet coupling function $\xi(\phi)$ to be a
particular choice, actually a dilaton-like coupling function like
\begin{equation}
\xi(\phi)=\xi_{0}e^{-\lambda\phi}, \label{cop}
\end{equation}
where $\xi_0$ and $\lambda$ are non-zero constants, $\phi$ is~not
constant, and in fact $\dot{\phi}\neq 0$. Although, for
convenience, one can reasonably reduce the number of free
parameters by introducing a parameter as $\alpha\equiv
4V_{0}\xi_{0}/3$.

Furthermore, in this model, the slow-roll parameters
(\ref{a14})-(\ref{a17}) are now functions of the inflaton scalar
field as
\begin{equation}
\epsilon_{1}\simeq\frac{2(4-\alpha\lambda e^{-\lambda\phi}\phi
^{5})}{\phi ^{2}}, \label{a27}
\end{equation}
\begin{equation}
\epsilon_{2}\simeq\frac{8-\alpha\lambda e^{-\lambda\phi}\phi
^{5}(\lambda\phi -3)}{\phi ^{2}}, \label{a28}
\end{equation}
\begin{equation}
\delta_{1}\simeq\frac{\alpha\lambda e^{-\lambda\phi}\phi
^{5}(4-\alpha\lambda e^{-\lambda\phi}\phi ^{5})}{\phi ^{2}},
\label{a29}
\end{equation}
\begin{equation}
\delta_{2}\simeq\frac{4(\lambda\phi -3)-2\alpha\lambda e^{-\lambda
\phi}\phi ^{5}(\lambda\phi -4)}{\phi ^{2}}. \label{a30}
\end{equation}
In turn, the tensor-to-scalar ratio and the scalar spectral index
(\ref{a24}) and (\ref{a25}) can also be obtained as a function of
the inflaton scalar field, i.e.,
\begin{equation}
r\simeq\frac{8(4-\alpha\lambda e^{-\lambda\phi}\phi ^{5})^{2}}{\phi ^{2}},
\label{a31}
\end{equation}
\begin{equation}
n_{\rm S}^{}-1\simeq\frac{-24+\alpha\lambda e^{-\lambda\phi}\phi
^{5}(2\lambda\phi -4)}{\phi ^{2}}. \label{a32}
\end{equation}

Before proceeding, investigating the proposed model in the absence
of the Gauss-Bonnet correction is also instructive. That is,
considering the general relativity limit with the quartic
potential when $\xi_0=0$ (that results in $\alpha=0$). In this
respect, the slow-roll parameters~(\ref{a27}) and~(\ref{a28})
reduce to the relevant ones in general
relativity\rlap,\footnote{In this case, the other two slow-roll
parameters~\eqref{a29}and~\eqref{a30} that were introduced due to
the presence of the Gauss-Bonnet term coupled to the inflaton
scalar field, no longer exist, i.e., by relations~\eqref{a8},
$\delta_n=0$ for all $n$.}\
 i.e.,
\begin{equation}\label{a27i}
\epsilon_{1}\simeq\frac{8}{\phi^{2}}\simeq\epsilon_{2}.
\end{equation}
In addition, the tensor-to-scalar ratio and the scalar spectral
index~(\ref{a31}) and~(\ref{a32}) are
\begin{equation}\label{a31i}
r\simeq\frac{128}{\phi^{2}}\qquad {\rm and}\qquad n_{\rm
S}^{}-1\simeq\frac{-24}{\phi^{2}}.
\end{equation}
Utilizing relation~\eqref{a12}, while using the value of
$\phi^2_{\rm end}\simeq 8$ from relation~\eqref{a27i} in the
absence of the Gauss-Bonnet correction, the relation
$\phi^2(N)\simeq 8(N+1)$ is obtained, and hence
relations~\eqref{a31i} read
\begin{equation}\label{a32i}
r\simeq\frac{16}{N+1}\qquad {\rm and}\qquad n_{\rm
S}^{}\simeq\frac{N-2}{N+1}.
\end{equation}
Thus, in this case, when we choose $N=70$, the best values for
inflationary observables can be achieved as
\begin{equation}\label{rn_sInGR}
r\simeq 0.2254\qquad {\rm and}\qquad n_{\rm S}^{}\simeq 0.9578,
\end{equation}
which roughly recover the predictions of the chaotic inflation
model~\cite{Liddle1999}. However, in the case of absence of the
Gauss-Bonnet correction, the value obtained above for $r$ is~not
consistent with the observable data.

Now, in the presence of the Gauss-Bonnet term, to proceed and
compare the obtained results with the observational data, the
functionality of $\phi$ should be specified and then inserted into
relations~(\ref{a31}) and (\ref{a32}). In this regard, a routine
procedure is to first acquire the inflaton scalar field as a
function of the $e$-folding number. Thereupon, by inserting the
best value of $e$-folding number (around $60-70$), the
inflationary observables can be computed straightforwardly. Hence,
we must solve relation~(\ref{a13}) by knowing the inflaton scalar
field at the end of inflation and then obtain the $e$-folding
number as a function of the inflaton scalar field. Next, by
deriving the inverse function, we can achieve the inflaton scalar
field as a function of the $e$-folding number.
\begin{figure*}[t!]
  \centering
  \includegraphics[width=0.7\textwidth]{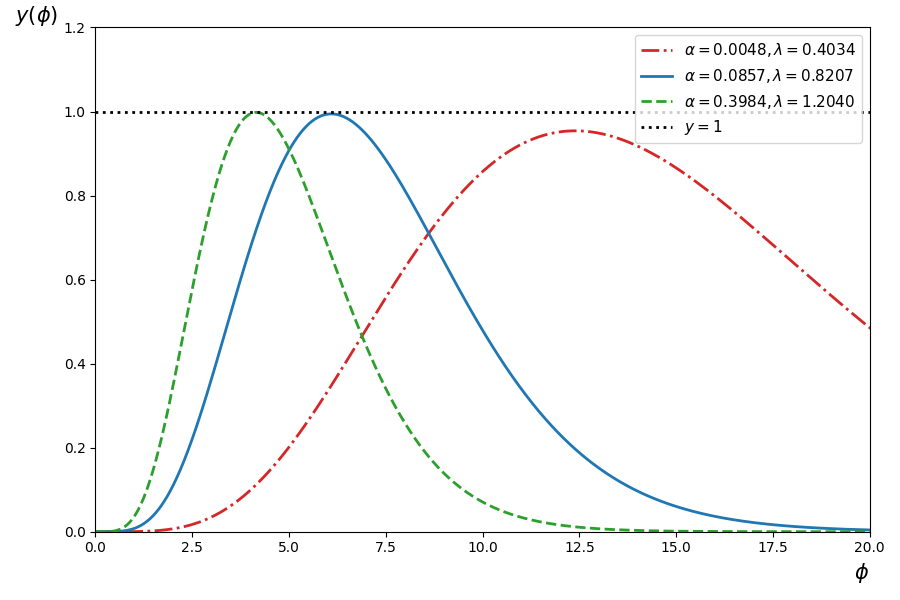}
  \caption{$y(\phi)= \alpha\lambda e^{-\lambda\phi}\phi^{5}/4$ versus $\phi$ for different
  values of $ \alpha $ and $\lambda$ while satisfying condition~(\ref{Condition2}).} \label{fa1}
  \includegraphics[width=0.7\textwidth]{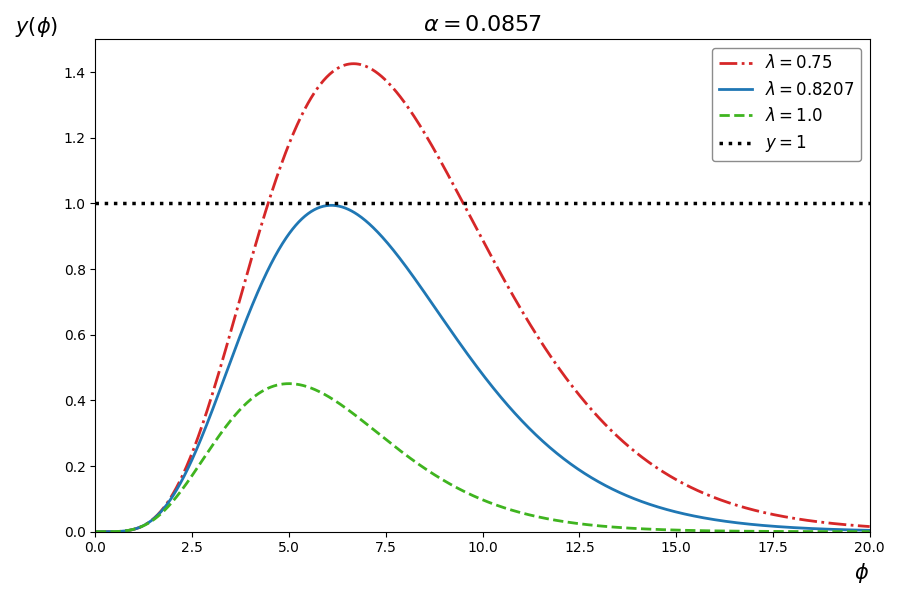}
  \caption{$y(\phi)= \alpha\lambda e^{-\lambda\phi}\phi ^{5}/4 $ versus $\phi$ for a fixed
  value of $ \alpha =0.0857 $ and different values of $ \lambda $.} \label{fa2}
\end{figure*}

In this regard, for the assumed potential and the coupling
function, relation~(\ref{a13}) is now
\begin{equation}
N(\phi)\simeq\int ^{\phi}_{\phi _{\rm end}}\frac{\phi
d\phi}{4-\alpha\lambda e^{-\lambda \phi}\phi^{5}}. \label{a33}
\end{equation}
However, even if the slow-roll approximations hold, this integral
is complicated and cannot be solved analytically. Hence, due to
its analytical complexities, to facilitate the study and
solvability of the system, certain approximations should be
applied. In this respect, to proceed, we first utilize the
well-known Taylor expansion
\begin{equation}
\frac{1}{1-x}=\sum _{n=0}^{\infty}x^{n}\qquad {\rm when\!:}\ \mid
x\mid <1, \label{a34}
\end{equation}
upon which, relation (\ref{a33}) reads
\begin{equation}
N(\phi)\simeq\sum
_{n=0}^{\infty}4^{-n-1}(\alpha\lambda)^{n}\int^{\phi}_{\phi_{\rm
end}}e^{-n\lambda \phi}\phi^{5n+1}d\phi, \label{a35}
\end{equation}
when
\begin{equation}\label{Condition1}
\vline\ \frac{\alpha\lambda}{4} e^{-\lambda\phi}\phi^{5}\
\vline<1.
\end{equation}
Then, within this condition, by analytically solving
relation~(\ref{a35}), we obtain
\begin{equation}
\begin{split}
    N(\phi)\simeq \frac{1}{8}(\phi^{2}-&\phi_{\rm end}^{2})+\sum _{n=1}^{\infty}4^{(-n-1)}(\alpha\lambda)^{n}(n\lambda)^{(-5n-2)} \\
     & \times\Big[ \gamma(5n+2;n\lambda\phi)-\!\gamma(5n+2;n\lambda\phi_{\rm end})\Big],\\
\end{split}
\label{a36}
\end{equation}
where $\gamma$ is the lower incomplete gamma function.
\begin{figure*}[t]
    \centering
    \includegraphics[width=0.95\textwidth]{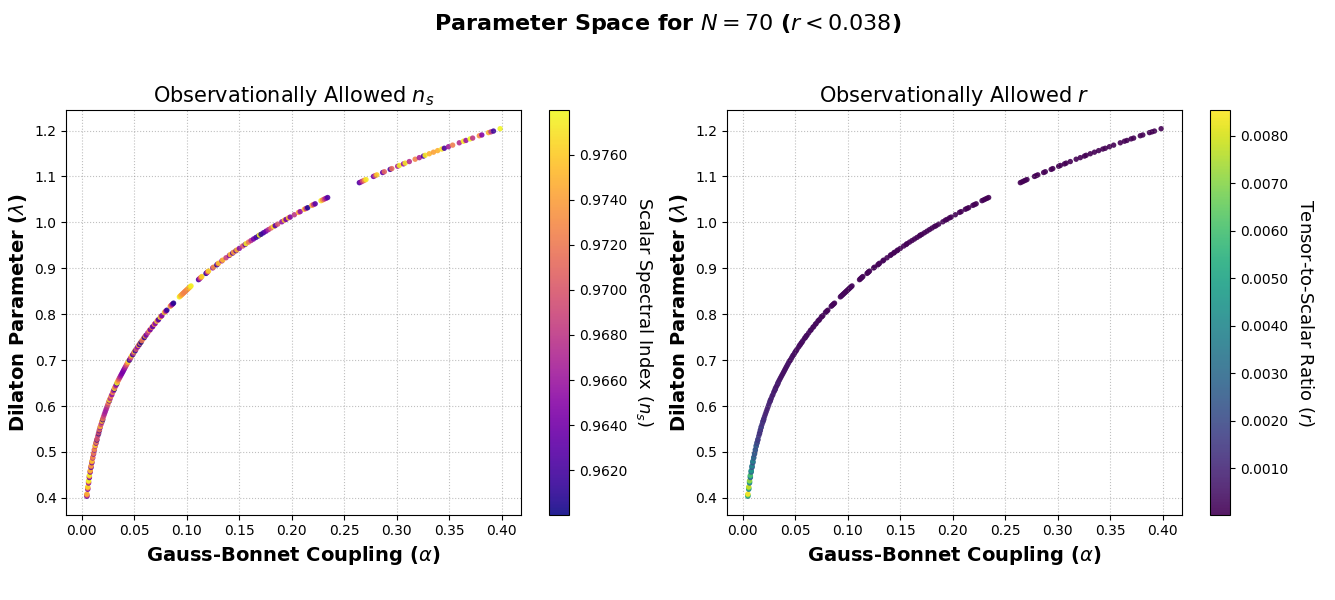}
    \caption{[color online] This figure shows the high-precision, observationally allowed parameter space in the $(\alpha, \lambda)$
    plane for $N=70$, obtained via the exact numerical integration of relation~\eqref{a33}. The left
    and right contour plots display the continuous variation of the scalar spectral index $n_{\rm S}^{}$ and the
    tensor-to-scalar ratio $r$, respectively. The obtained (colored) regions strictly satisfy both the
    mathematical condition~\eqref{Condition2} and the stringent observational
    constraints within the ranges $0.96 < n_{\rm S}^{} < 0.978$ and $r < 0.038$.}
    \label{fa3}
\end{figure*}
\begin{table*}[t]
\centering \caption{The analytical approximation results for the
selected benchmark points at $N(\phi_*)=70$. The values
demonstrate a viable spectrum, ranging from weak to strong
Gauss-Bonnet coupling effects, while strictly satisfying the
observational constraints on $n_{\rm S}^{}$ and $r$.}
\vspace{0.2cm} \label{tab1}
\begin{tabular}{cc|cccc|cc|ccc|c}
\hline\hline
 $\boldsymbol{\alpha}$  &
$\boldsymbol{\lambda}$ & $\boldsymbol{\epsilon_1}$ &
$\boldsymbol{\epsilon_2}$ & $\boldsymbol{\delta_1}$ & $
\boldsymbol{\delta_2}$ & $\boldsymbol{\phi_{\rm end} }$ &
$\boldsymbol{\phi_* }$ & $\boldsymbol{n_{\rm S}^{}}$
& $\boldsymbol{n_{\rm T}^{}}$ & $\boldsymbol{r}$ & $\boldsymbol{\mathcal{R}(\phi)}\%$ \\
\hline \!\!\!\!\!\!\! 0.0048
  \!\!\!\!&\!\!\!\! 0.4034
  \!\!\!\!&\!\!\!\! 0.00399
  \!\!\!\!&\!\!\!\! 0.0164
  \!\!\!\!&\!\!\!\! 0.00746
  \!\!\!\!&\!\!\!\! 0.0156
  \!\!\!\!&\!\!\!\! 2.79059
  \!\!\!\!&\!\!\!\! 11.33754
  \!\!\!\!&\!\!\!\! 0.96320
  \!\!\!\!&\!\!\!\! $-0.00798$
  \!\!\!\!&\!\!\!\! $0.409\times 10^{-2}$
  \!\!\!\!&\!\!\!\! $0.732\times 10^{-43}$ \\
 \!\!\!\!\!\!\! 0.0215
  \!\!\!\!&\!\!\!\! 0.5823
  \!\!\!\!&\!\!\!\! 0.00249
  \!\!\!\!&\!\!\!\! 0.0158
  \!\!\!\!&\!\!\!\! 0.00488
  \!\!\!\!&\!\!\!\! 0.0155
  \!\!\!\!&\!\!\!\! 2.69435
  \!\!\!\!&\!\!\!\! 8.19379
  \!\!\!\!&\!\!\!\! 0.96584
  \!\!\!\!&\!\!\!\! $-0.00498$
  \!\!\!\!&\!\!\!\! $0.833\times 10^{-3}$
  \!\!\!\!&\!\!\!\! $0.517 \times 10^{-13}$ \\
 \!\!\!\!\!\!\! 0.0572
  \!\!\!\!&\!\!\!\! 0.7424
  \!\!\!\!&\!\!\!\! 0.00193
  \!\!\!\!&\!\!\!\! 0.0134
  \!\!\!\!&\!\!\!\! 0.00381
  \!\!\!\!&\!\!\!\! 0.0133
  \!\!\!\!&\!\!\!\! 2.56740
  \!\!\!\!&\!\!\!\! 6.56611
  \!\!\!\!&\!\!\!\! 0.97121
  \!\!\!\!&\!\!\!\! $-0.00386$
  \!\!\!\!&\!\!\!\! $0.32\times 10^{-3}$
  \!\!\!\!&\!\!\!\! $0.101\times 10^{-5}$\\
 \!\!\!\!\!\!\! 0.0857
  \!\!\!\!&\!\!\!\! 0.8207
  \!\!\!\!&\!\!\!\! 0.00157
  \!\!\!\!&\!\!\!\! 0.0146
  \!\!\!\!&\!\!\!\! 0.00313
  \!\!\!\!&\!\!\!\! 0.0145
  \!\!\!\!&\!\!\!\! 2.49783
  \!\!\!\!&\!\!\!\! 5.95038
  \!\!\!\!&\!\!\!\! 0.96913
  \!\!\!\!&\!\!\!\! $-0.00314$
  \!\!\!\!&\!\!\!\! $0.176\times 10^{-3}$
  \!\!\!\!&\!\!\!\! $0.223 \times 10^{-3}$ \\
 \!\!\!\!\!\!\! 0.1207
  \!\!\!\!&\!\!\!\! 0.8940
  \!\!\!\!&\!\!\!\! 0.00156
  \!\!\!\!&\!\!\!\! 0.0114
  \!\!\!\!&\!\!\!\! 0.00310
  \!\!\!\!&\!\!\!\! 0.0114
  \!\!\!\!&\!\!\!\! 2.43170
  \!\!\!\!&\!\!\!\! 5.50862
  \!\!\!\!&\!\!\!\! 0.97559
  \!\!\!\!&\!\!\!\! $-0.00312$
  \!\!\!\!&\!\!\!\! $0.147\times 10^{-3}$
  \!\!\!\!&\!\!\!\! $0.150 \times 10^{-2}$\\
 \!\!\!\!\!\!\! 0.2341
  \!\!\!\!&\!\!\!\! 1.0541
  \!\!\!\!&\!\!\!\! 0.00097
  \!\!\!\!&\!\!\!\! 0.0169
  \!\!\!\!&\!\!\!\! 0.00193
  \!\!\!\!&\!\!\!\! 0.0169
  \!\!\!\!&\!\!\!\! 2.28679
  \!\!\!\!&\!\!\!\! 4.66083
  \!\!\!\!&\!\!\!\! 0.96514
  \!\!\!\!&\!\!\!\! $-0.00194$
  \!\!\!\!&\!\!\!\! $0.41\times 10^{-4}$ & 0.281 \\
 \!\!\!\!\!\!\! 0.3984
  \!\!\!\!&\!\!\!\! 1.2040
  \!\!\!\!&\!\!\!\! 0.00100
  \!\!\!\!&\!\!\!\! 0.0095
  \!\!\!\!&\!\!\!\! 0.00200
  \!\!\!\!&\!\!\!\! 0.0095
  \!\!\!\!&\!\!\!\! 2.15829
  \!\!\!\!&\!\!\!\! 4.12281
  \!\!\!\!&\!\!\!\! 0.98002
  \!\!\!\!&\!\!\!\! $-0.00200$
  \!\!\!\!&\!\!\!\! $0.34 \times 10^{-4}$& 0.876 \\
\hline\hline
\end{tabular}
\end{table*}

Meanwhile, let us examine condition (\ref{Condition1}) carefully.
The functionality of $y(\phi)\equiv\alpha\lambda
e^{-\lambda\phi}\phi ^{5}/4$ indicates that its value is always
positive and has a maximum for every positive value of $\alpha$
and $\lambda$, for instance, see
Fig.~(\ref{fa1})\rlap.\footnote{The sample values mentioned for
$\alpha$ and $\lambda$ are among the acceptable ones that we will
acquire in Fig.~(\ref{fa3}).}\
 Furthermore, the value of $\phi$ for the
peak of the curve of $y(\phi)$ is $\phi|_{_{\rm P}}=5/\lambda$,
and hence $y_{\rm max}(\phi|_{_{\rm
P}})=e^{-5}5^5\alpha\lambda^{-4}/4$. Thus, by
condition~(\ref{Condition1}), we obtain the constraint
\begin{equation}\label{Condition2}
\alpha\lambda^{-4}<4\,e^55^{-5}\sim 0.19
\end{equation}
on positive values of $\alpha$ and $\lambda$. Alternatively, for
any fixed value of $\alpha$, there is a lower-bound on $\lambda$
due to condition~(\ref{Condition2}). For instance, see
Fig.~(\ref{fa2}) for $ \alpha=0.0857 $, where $ \lambda $ can have
any value in the interval $ [0.8207,\infty) $. Moreover,
Fig.~(\ref{fa2}) illustrates that for a fixed value of $\alpha$,
increasing $ \lambda$ from its lower-bound reduces the peak value
of $y(\phi)$, making it more consistent with the convergence
condition~(\ref{Condition1}). Besides, it would be constructive to
have the slow-roll parameters (\ref{a27})-(\ref{a30}) and the
inflationary observables (\ref{a31}) and (\ref{a32}) in terms of
the function $y(\phi)$ as
\begin{equation}\label{Newa27.1}
\epsilon_1\simeq 8(1-y)/\phi^2,
\end{equation}
\begin{equation}\label{Newa27.2}
\delta_1\simeq 16y(1-y)/\phi^2,
\end{equation}
\begin{equation}\label{Newa29}
\epsilon_2\simeq 4\left[2-y(\lambda\phi-3)\right]/\phi^2,
\end{equation}
\begin{equation}\label{Newa30}
\delta_{2}\simeq
4\left[\lambda\phi-3-2y(\lambda\phi-4)\right]/\phi^2,
\end{equation}
\begin{equation}\label{Newa31}
r\simeq 128(1-y)^2/\phi^2,
\end{equation}
\begin{equation}\label{Newa32}
n_{\rm S}^{}-1\simeq 4\left[-6+y(2\lambda\phi-4)\right]/\phi^2.
\end{equation}
Here, by inverting the defined function $y(\phi)$, we have
$\phi(y)=-5\,W[-\lambda (4y/\alpha\lambda)^{1/5}/5]/\lambda$,
where $W$ denotes the Lambert (product logarithm) function.

At this stage, since the main purpose of this work is to implement
and investigate the impact of the proposed modified gravity model
on the inflationary observables, it is necessary to choose the
model parameters such that they yield acceptable values for the
$e$-folding number as well as the other inflationary observables.
Hence, to incorporate such a constraint and ensure the correctness
of the values used for the model parameters, we have drawn
Fig.~\eqref{fa3}. Indeed, to identify the allowed range of the
model parameters, we have performed a scan of the $(\alpha,
\lambda)$ plane. In this regard, although the $1500$-term series
expansion of relation~\eqref{a36}\footnote{The series expressed in
relation~\eqref{a36} exhibits very slow convergence.}\
 provides high-precision for individual points, scanning
the entire parameter space is computationally intensive due to the
requirement to evaluate numerous lower incomplete gamma functions.
Alternatively, for an efficient mapping of the allowed range, we
have performed a direct, exact numerical integration of
relation~\eqref{a33}. This approach is significantly more
efficient and, as demonstrated later, introduces infinitesimal
discrepancy compared to the aforementioned $1500$-term series.
Accordingly, in Fig.~\eqref{fa3}, for each point $(\alpha,
\lambda)$ that satisfies condition~\eqref{Condition2}, we first
numerically determined $\phi_{\rm end}$ from relation~\eqref{a27}
at the end of inflation, and subsequently have calculated
$\phi_{*}$ via direct integration of relation~\eqref{a33} subject
to $N(\phi_{*}) = 70$. The results of this scan are presented in
Fig.~\eqref{fa3}, showing the allowed range for the $\alpha$
parameter as $[0.0048, 0.3984]$ and for the $\lambda$ parameter as
$[0.4034, 1.2040]$. However, the obtained (colored) regions for
the inflationary parameters indicate compatibility with the latest
Planck, BICEP/Keck, and ACT DR6 data within the parameter ranges
$0.96 < n_{\mathrm{S}} < 0.978$ and $r < 0.038$ at $N = 70$.

Once the viable ranges are identified through this efficient
numerical scan of the $(\alpha, \lambda)$ plane, we have selected
seven representative benchmark points that span the allowed range
for $\alpha$ and $\lambda$. For these selected points, we have
performed the full calculation using the $1500$-term series
expansion of relation~\eqref{a36} to obtain high-precision values
for the inflationary observables. Table~\ref{tab1} presents these
points along with the corresponding values of $\phi_{*}$, $n_{\rm
S}^{}$, $n_{\rm T}^{}$, and $r$, all computed using the
$1500$-term series expansion of relation~\eqref{a36}. For
completeness, additional quantities--including $\phi_{\rm end}$,
the slow-roll parameters, and the relative discrepancy (defined
below)--are also listed in Table~\ref{tab1}.
\begin{figure*}[t!]
  \centering
  \includegraphics[width=0.7\textwidth]{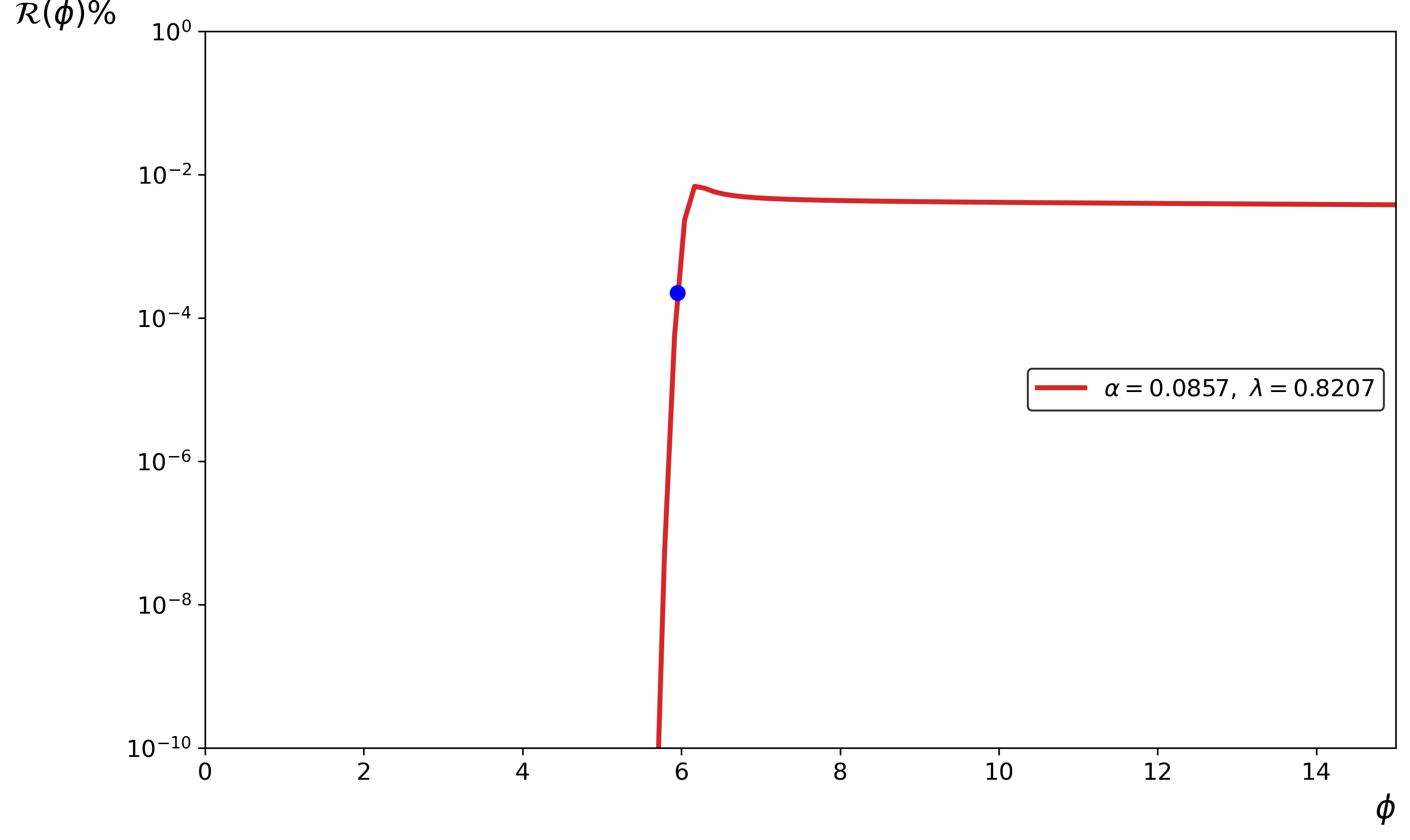}
  \caption{$\mathcal{R}$ versus $\phi$ for the values $ \alpha =0.0857 $ and $ \lambda
  =0.8207$, where the
  blue point shows $ N = 70$. }\label{fa4}
\end{figure*}

On the other hand, to validate the accuracy of this numerical scan
and confirm that the use of exact integration for the parameter
space exploration does~not compromise precision, we have performed
a detailed error analysis. In fact, for the selected benchmark
points, we have compared the $1500$-term series expansion of
relation~\eqref{a36} with the exact numerical integration of
relation~\eqref{a33}. In this regard, the relative discrepancy,
defined as
\begin{equation}
\mathcal{R}(\phi) = \left| \frac{N_{\mathrm{exact}}(\phi) -
N_{\mathrm{series}}(\phi)}{N_{\mathrm{exact}}(\phi)} \right|,
\end{equation}
is consistently found to be very low at horizon crossing
$\phi_{*}$. In particular, for the smallest parameter $\alpha =
0.0048$, the relative discrepancy is less than $10^{-43}\%$, while
for the largest parameter $\alpha = 0.3984$, it remains below
$0.9\%$. The results of relative discrepancy for all the selected
benchmark points are given in Table~\ref{tab1}. Also, as a sample
case, we have chosen the parameters $ \alpha=0.0857$ and $
\lambda=0.8207$, and have plotted its relative discrepancy versus
$\phi$ in Fig.~\eqref{fa4}. The remarkably small relative
discrepancy across the parameter space confirms two crucial
points. First, the Taylor expansion, when carried out to a
sufficiently high-order, is mathematically consistent within its
radius of convergence. Second, the two-stage approach, which uses
the exact integration for efficient scanning and the $1500$-term
series for final precision, is both computationally practical and
mathematically robust, introducing no~significant loss of
accuracy.

Having established the validity of the employed methodology, we
now proceed to examine the inflationary observables with high
precision. The results presented in Table~\ref{tab1} demonstrate
that the Gauss-Bonnet correction can successfully reduce the
tensor-to-scalar ratio $r$ to values well within the observational
bounds, while maintaining the scalar spectral index
$n_{\mathrm{S}}$ within the range favored by the Planck and ACT
DR6 data. Notably, for all the selected benchmark points, $r$ is
significantly smaller than the prediction of the quartic potential
in general relativity, highlighting the crucial role of the
Gauss-Bonnet coupling in suppressing tensor perturbations. In
fact, the sufficiently low value obtained for the tensor-to-scalar
ratio lies comfortably below the experimental limits on primordial
gravitational waves. However, in the absence of the Gauss-Bonnet
term, where the model effectively reduces to general relativity,
its obtained value (relation~\eqref{rn_sInGR}) is far from the
observable data. Consequently, these new consistent values clearly
demonstrate the significant impact of incorporating the
Gauss-Bonnet correction, especially when using the chosen coupling
function. To proceed, let us focus on one of the intermediate
benchmark points, specifically $ \alpha = 0.0857$ and $ \lambda =
0.8207$, and inspect further calculations for this sample point.
Accordingly, the slow-roll parameters during inflation for this
point are
\begin{eqnarray}
 && \epsilon_1\simeq 1.57\times 10^{-3},\qquad\,\ \epsilon_2\simeq 1.46\times 10^{-2},\cr
 && \delta_1\simeq 3.13\times 10^{-3},\qquad\,\ \delta_2\simeq 1.45\times 10^{-2},
\end{eqnarray}
respectively, that are well below one as
expected\rlap.\footnote{Since we have used numerical analysis
methods, one needs to ensure these smallnesses. }\
 In turn,
utilizing relation~\eqref{a26}, the tensor spectral index is
\begin{equation}
n_{\rm T}^{}\simeq -3.14\times 10^{-3}.
\end{equation}
These results for all the selected benchmark points are presented
in Table~\ref{tab1}. Moreover, for that intermediate benchmark
point, the value of the $y(\phi)$ function is approximately
$0.99297$, which is less than one, as expected.
\begin{figure*}[t!]
  \centering
  \includegraphics[width=0.7\textwidth]{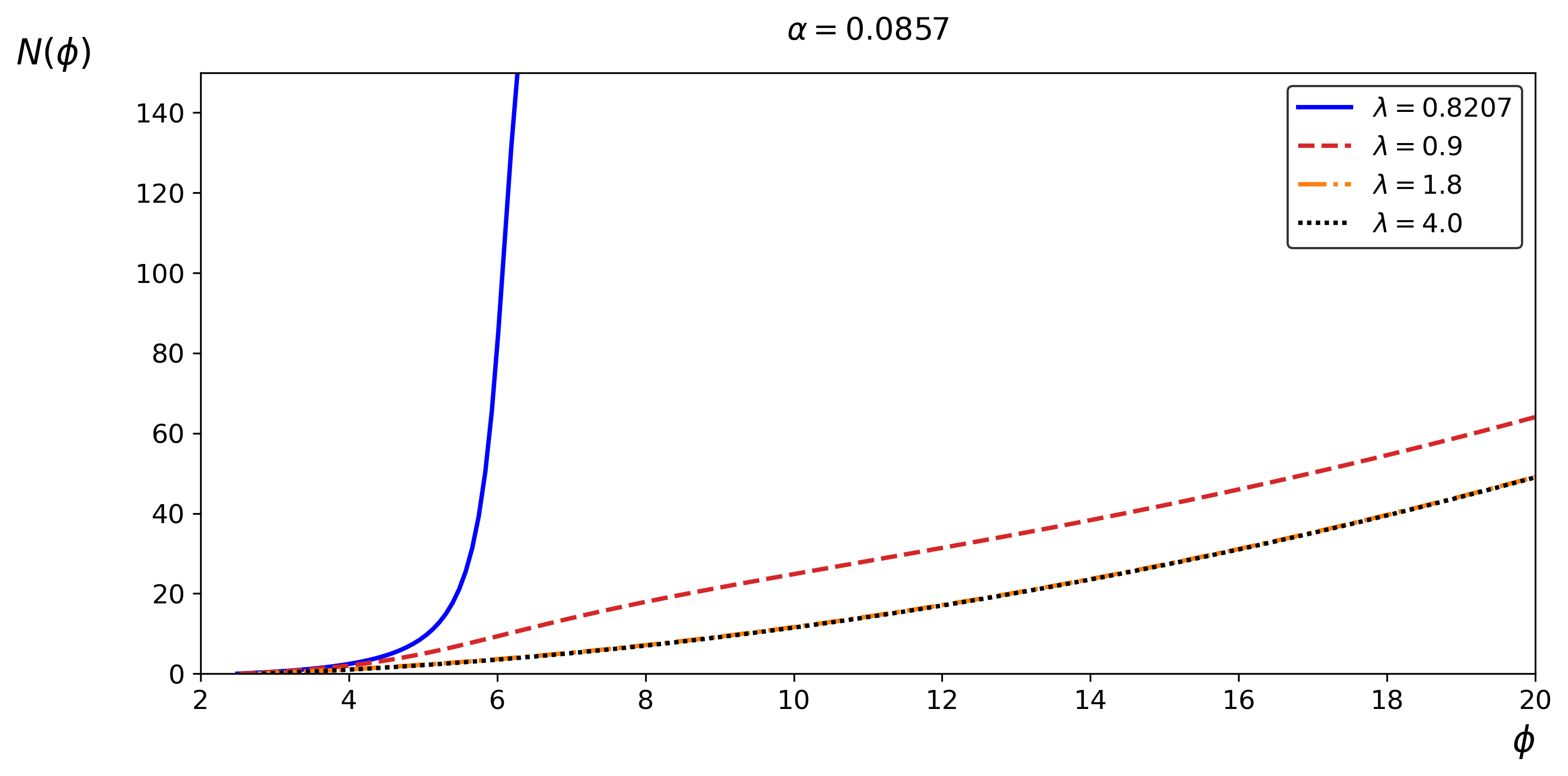}
  \caption{$N(\phi)$ versus $\phi$ for a fixed
  value of $ \alpha =0.0857 $ and different values of $ \lambda $. }\label{fa5}
 \includegraphics[width=0.7\textwidth]{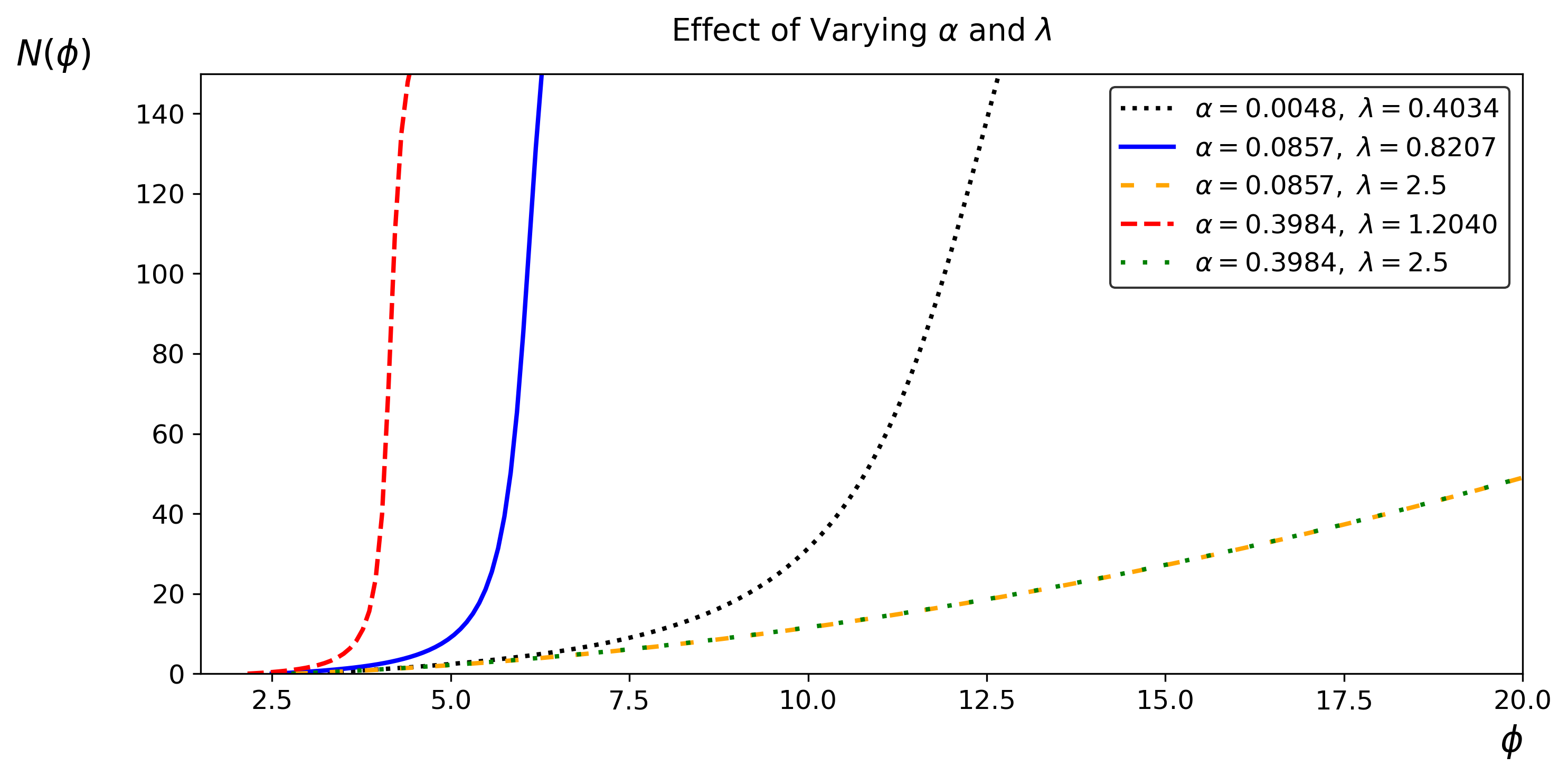}
  \caption{$N(\phi)$ versus $\phi$ for different values of $ \alpha $ with the corresponding
  minimum value of $ \lambda $ in each case. }\label{fa6}
\end{figure*}

Also, to insure acceptable values for the $e$-folding number
within the allowed range of the model parameters, we have plotted
the $e$-folding function~(\ref{a36}) versus $\phi$ in
figures~(\ref{fa5}) and~(\ref{fa6}). However, the plots were
performed using a series expansion of approximately $1500$ terms,
for different values of the model parameters obtained from
Fig.~(\ref{fa3}), alongside their corresponding $\phi_{\rm end}$
values. Fig.~\eqref{fa5} illustrates that for a fixed value of
$\alpha$, and contingent upon the value of the scalar field, there
is also an upper-bound for the corresponding $\lambda$ parameter,
beyond which acceptable values for the $e$-folding number may
no~longer be achieved. Furthermore, Fig.~\eqref{fa6} displays the
evolution of e-folding values for various values of the $\alpha$
parameter. This also highlights the sensitivity of the
inflationary dynamics and the impact of the Gauss-Bonnet term with
the chosen coupling function on its strength. This observation
makes it possible to determine how the field excursion required
for $N \simeq 70$ e-folds scales with the magnitude of the
coupling parameter $\lambda$. Indeed, the exact numerical
integration reveals two significant findings regarding the
$\lambda$ parameter. First, increasing $\alpha$ shifts the
$N(\phi)$ trajectories toward larger values. Physically, a
stronger Gauss-Bonnet coupling modifies the effective friction
experienced by the inflaton field. Consequently, to accumulate the
required $70$ e-folds that are necessary to resolve the standard
cosmological problems, the field must commence its slow-roll from
a higher initial value $\phi_*$. This confirms that enhancing the
coupling parameter $\lambda$ naturally pushes the dynamics further
into the large-field inflationary regime. Second, a deeper
inspection of the curves in Fig.~\eqref{fa6} reveals a distinctive
asymptotic behavior, or a saturation limit. In fact, the
denominator of the exact integrand of the e-folding number in
relation~\eqref{a33} accounts for this saturation. That is, for
sufficiently large values of the field $\phi$, the exponential
suppression factor $e^{-\lambda \phi}$ heavily dominates over both
the polynomial $\phi^5$ and the $\alpha$ parameter. Consequently,
as $e^{-\lambda \phi} \to 0$, the second term in the denominator
effectively vanishes, and the denominator asymptotically becomes
the constant value of $4$. Hence, in this limit, the inflationary
dynamics loses its sensitivity to the Gauss-Bonnet modification
and elegantly converges back to the standard general relativity
scenario, yielding $N(\phi) \simeq \phi^2/8-1$. This described
mechanism explains why the $N(\phi)$ curves saturate beyond a
certain threshold in the field excursion, and the theory
effectively reduces to the uncoupled standard inflation. It is
worth noting that, by utilizing the values of the selected
benchmark parameters and the two approximate values of $V_0$
mentioned in Footnote~$3$, the estimated value of the coefficient
of the coupling function to the Gauss-Bonnet term in
action~\eqref{a1}, i.e., $\xi_0$, is approximately of the order of
${\cal O}(10^{-1})-{\cal O}(10^1)$ or ${\cal O}(10^{11})-{\cal
O}(10^{13})$, respectively.
\begin{figure*}[t!]
  \centering
  \includegraphics[width=0.7\textwidth]{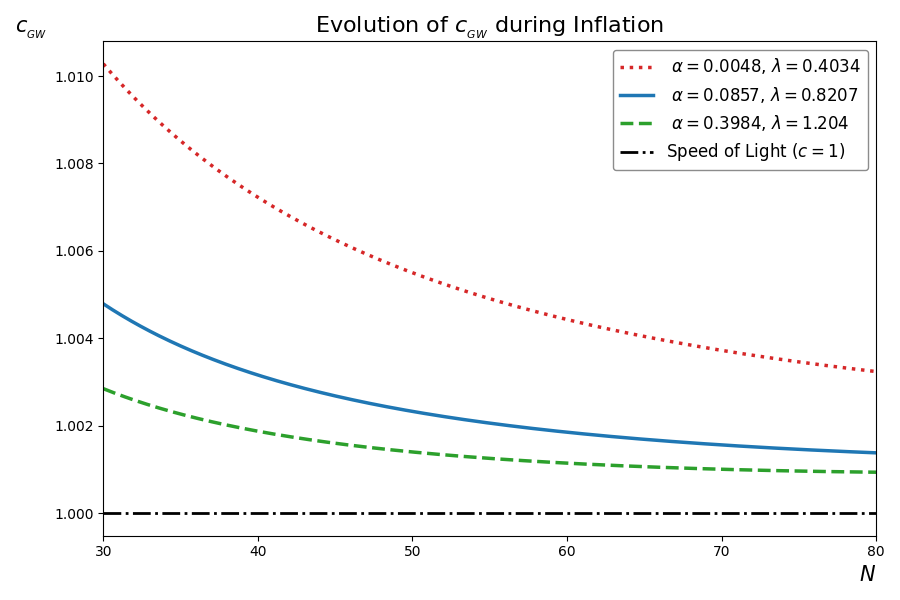}
  \caption{The speed of gravitational waves $c_{_{\rm GW}}$ versus $N$ for
  a few values of $\alpha$ and $\lambda$ during inflation.}\label{fa7}
\end{figure*}
\begin{table*}[t]
\centering \caption{The speed of gravitational waves at the end of
inflation for the values of $\alpha$ and $\lambda$.}
\vspace{0.2cm} \label{tab2}
\begin{tabular}{|c|c|c|c|}
\hline\hline
 $\boldsymbol{\alpha}$\qquad  &
$\boldsymbol{\lambda}$\qquad  & $\boldsymbol{\phi_{\rm end}
}$\qquad
& $\boldsymbol{c_{_{\rm GW}}\!\!\mid_{_{\phi_{\rm end}}}}$ \quad\\
 \hline
0.0048  & 0.4034 & 2.79059 & 1.0262 \\
0.0215 & 0.5823 & 2.69435 & 1.0886 \\
0.0572 & 0.7424 & 2.56740 & 1.1628 \\
0.0857 & 0.8207 & 2.49783 & 1.2001 \\
0.1207 & 0.8940 & 2.43170 & 1.2336 \\
0.2341 & 1.0541 & 2.28679 & 1.3010 \\
0.3984 & 1.2040 & 2.15829 & 1.3548\\
\hline\hline
\end{tabular}
\end{table*}

At this stage, let us examine the speed of gravitational waves
during inflation. In this regard, relation~\eqref{a21} shows the
propagation speed of tensor perturbation modes (including
gravitational waves) in the presence of the Gauss-Bonnet
correction. During inflation, utilizing the approximations in
relation~\eqref{Conditions3,4} into relation~\eqref{a21}, we
obtain
\begin{equation}\label{gw1}
c_{_{\rm GW}}^{2}\! \simeq\!
1\!+\frac{4H\dot{\xi}}{1\!-\!4H\dot{\xi}}\!\simeq\!
1\!+4H\dot{\xi}(1\!+4H\dot{\xi})\!\simeq\! 1\!+4H\dot{\xi}\!=\!
1\!+\delta_{1},
\end{equation}
where in the last equality we have used relation~\eqref{a8}. Thus,
under the approximate conditions used for inflation, if the time
derivative of the coupling function is positive, the speed of
gravitational waves during inflation will be greater than the
speed of light. Now, using relation~\eqref{a29}, while writing
$\phi(N)$, relation~\eqref{gw1} gives the speed of gravitational
waves during inflation in terms of the $e$-folding number as
\begin{equation}
c_{_{\rm GW}}^{2}(N)\! \simeq\! 1\! +\alpha\lambda
e^{-\lambda\phi(N)}\phi^{3}(N)\left[4\!-\!\alpha\lambda
e^{-\lambda\phi(N)}\phi^{5}(N)\right].
 \label{gw2}
\end{equation}
Then, we have plotted the speed of gravitational waves within
inflation for three benchmark points in Fig.~\eqref{fa7}. However,
since relation~(\ref{a36}) is intricate and cannot be inverted
analytically, we utilized Fig.~\eqref{fa6} to obtain $\phi(N)$
values for each pair of $(\alpha, \lambda)$ values through
numerical analysis. Thereafter, the results were inserted in
relation~(\ref{gw2}). Fig.~\eqref{fa7} shows that the speed of
gravitational waves during inflation is greater than one (i.e.,
the speed of light) but very close to it, and its very slight
difference from one is due to the presence of the Gauss-Bonnet
correction. Hence, the contribution of the Gauss-Bonnet term to
inflationary spectra in this Higgs inflation-like model during
inflation is almost negligible, although not~absolutely zero. Of
course, if $\dot{\xi}=0$ in relation~\eqref{gw1}, this speed will
be exactly equal to one and the Gauss-Bonnet correction has
no~effect at all. In fact, in such a situation, since $\xi$ will
be a constant value, the Gauss-Bonnet term will have no~effect on
the field equations in four dimensions.

Furthermore, at the end of inflation, utilizing
$\epsilon_1\!\!\mid_{\phi_{\rm end}}^{}\simeq~\!\!1$ from
relation~\eqref{a27} in relation~\eqref{a29}, we arrive at
\begin{equation}\label{a29AtEnd}
\delta_{1}\!\!\mid_{\phi_{\rm end}}^{}\simeq 2-\frac{\phi_{\rm
end}^2}{4}.
\end{equation}
Then, by substituting it into relation~\eqref{gw1}, we obtain the
speed of gravitational waves at the end of inflation as
\begin{equation}\label{gwend}
c_{_{\rm GW}}^{2}\!\!\mid_{\phi_{\rm end}}^{} \simeq
3-\frac{\phi_{\rm end}^{2}}{4}.
\end{equation}
In Table~\eqref{tab2}, we have presented its values for the
selected benchmark points $\alpha$ and $\lambda$. Although the
Gauss-Bonnet term exerts a significant influence on the underlying
physics of the system at the end of the inflationary era, its very
slight difference from one justifies that if gravitons are
massless after the inflationary era, compatibility can be achieved
by equating the speed of gravitational waves to the speed of light
or infinitesimally close to it.

\section{Conclusions}
\indent

We have considered the Gauss-Bonnet term coupled with the Higgs
scalar field added to the Einstein-Hilbert action. Then, we have
derived the equations of motion and, employing the FLRW metric,
the modified Friedmann equations. Thereafter, we have approximated
these equations during inflation by considering the appropriate
slow-roll parameters. Moreover, the proposed model includes the
Higgs potential and the coupling coefficient as a dilaton-like
coupling.

Under these assumptions, we encountered calculations to obtain the
tensor-to-scalar ratio and the scalar spectral index that are
rather complicated and cannot be easily solved analytically.
Hence, further approximations have been needed in order to derive
the inflationary phenomenology. In this regard, in the first
stage, we did~not proceed directly with numerical analysis
methods. Instead, we found that it productive to first employ a
well-known Taylor expansion, after which we could solve the
$e$-folding integral analytically. Meanwhile, this procedure has
caused a constraint on the model parameters. However, to obtain
acceptable values for the $e$-folding function, we determined it
through numerical analysis methods for a certain range of derived
values for the model parameters.

In the process of this work, several plots have been drawn, and at
the same time quasi-numerical analysis methods have been employed.
Subsequently, by considering the aforementioned constraint on the
model parameters, we have performed a scan of their plane. Within
this scan, we scrutinized the parameter space to attain the
optimal contour plots for the tensor-to-scalar ratio and the
scalar spectral index, ensuring consistency with the latest
observational data, particularly from ACT DR6. In the efficient
numerical scan, we have specified appropriate ranges for the model
parameters that yield those good results for the inflationary
observables. Furthermore, within these specified ranges, we also
specified the $e$-folding values that fall within the acceptable
range of observational data. It is noteworthy that a key aspect of
this work, crucial for achieving appropriate results for the
inflationary observables, lies in the adopted functional forms for
the potential and coupling functions.

Also, in the absence of the Gauss-Bonnet term, we find that the
inflationary observables are roughly the same as the predictions
of the chaotic inflation model. However, in this case, the value
obtained for the tensor-to-scalar ratio is~not consistent with the
observable data, while the very acceptable values obtained for it
in the presence of the Gauss-Bonnet term indicate the significant
effect of including the Gauss-Bonnet correction. In addition, the
results of the model show that the speed of gravitational waves
during inflation is greater than one (i.e., the speed of light)
but very close to it. This slight difference from one is due to
the presence of the Gauss-Bonnet correction.

\section*{ACKNOWLEDGMENTS}
\indent

The authors thank the Deputy for Research and Technology of Shahid
Beheshti University.


\end{document}